%% file: 0_main.tex
\documentclass[sigconf]{acmart}
\AtBeginDocument{%
  \providecommand\BibTeX{{%
    \normalfont B\kern-0.5em{\scshape i\kern-0.25em b}\kern-0.8em\TeX}}}

\setcopyright{acmcopyright}
\copyrightyear{2024}
\acmYear{2024}
\acmDOI{XXXXXXX.XXXXXXX}

\acmConference[WSDM 2024]{The 17th ACM International Conference on Web Search and Data Mining}{March 4th-8th, 2024}{Merida,Yucatan,Mexico}
\acmPrice{15.00}
\acmISBN{978-1-4503-XXXX-X/18/06}

\usepackage{xspace}
\usepackage{amsmath,amsfonts}  % amssymb
\usepackage{amsthm}
\usepackage{multirow}
\usepackage{listings}
\usepackage{hyperref}
\usepackage{tabu}
\usepackage{xcolor}         % colors

\hypersetup{
    colorlinks=true,
    urlcolor=cyan
    }

\copyrightyear{2024}
\acmYear{2024}
\setcopyright{acmlicensed}\acmConference[WSDM '24]{Proceedings of the 17th ACM International Conference on Web Search and Data Mining}{March 4--8, 2024}{Merida, Mexico}
\acmBooktitle{Proceedings of the 17th ACM International Conference on Web Search and Data Mining (WSDM '24), March 4--8, 2024, Merida, Mexico}
\acmPrice{15.00}
\acmDOI{10.1145/3616855.3635694}
\acmISBN{979-8-4007-0371-3/24/03}

\input{000macros}

\begin{document}

% \title{TGX: A Package for Temporal Graph Analysis and Visualization}
\title{Temporal Graph Analysis with TGX }
% Temporal Graph Analysis with TGX 

\input{000authors}


\begin{abstract}

\input{00abstract}

\end{abstract}


% \ccsdesc[500]{Software and its engineering~Software creation and management}



\maketitle

\input{01introduction}
\input{04Preliminaries}
\input{02Features}
\input{05Conclusion}
\input{08Ack}

\bibliographystyle{ACM-Reference-Format}
\bibliography{09References}

\end{document}

%% file: 000macros.tex
\newcommand{\tgx}{TGX\xspace}
\NewDocumentCommand{\codeword}{v}{%
\texttt{\textcolor{blue}{#1}}%
}
\newcommand{\mat}[1]{\ensuremath{\mathbf{#1}}}

%% file: 000authors.tex
% \author{Razieh Shirzadkhani, Shenyang Huang, Elahe Kooshafar, Reihaneh Rabbany, Farimah Poursafaei}
% \affiliation{%
%   \institution{McGill University, School of Computer Science \& Mila --- Quebec AI Institute}
%   % \city{Montreal, QC}
%   % \country{Canada}
% }
% \email{{ razieh.shirzadkhani, huangshe, elahe.kooshafar, reihaneh.rabbany, farimah.poursafaei }@mila.quebec}

\author{Razieh Shirzadkhani}
\affiliation{%
  \institution{McGill University \& Mila}
  \country{Montreal, QC, Canada}
}
\email{razieh.shirzadkhani@mail.mcgill.ca}

\author{Shenyang Huang}
\affiliation{%
  \institution{McGill University \& Mila}
  \country{Montreal, QC, Canada}
}
\email{shenyang.huang@mail.mcgill.ca}

\author{Elahe Kooshafar}
\affiliation{%
  \institution{McGill University \& Mila}
  \country{Montreal, QC, Canada}
}
\email{elahe.kooshafar@mila.quebec}

\author{Reihaneh Rabbany}
\affiliation{%
  \institution{McGill University \& Mila}
  \country{Montreal, QC, Canada}
}
\email{rrabba@cs.mcgill.ca}

\author{Farimah Poursafaei}
\affiliation{%
  \institution{McGill University \& Mila}
  \country{Montreal, QC, Canada}
}
\email{farimah.poursafaei@mila.quebec}

\renewcommand{\shortauthors}{Razieh Shirzadkhani, Shenyang Huang, Elahe Kooshafar, Reihaneh Rabbany, \& Farimah Poursafaei}

% \author{Razieh Shirzadkhani, Elahe Kooshafar, Farimah Poursafaei, Shenyang Huang, Reihaneh Rabbany}
% \affiliation{%
%   \institution{McGill University \& Mila, Montreal, QC, Canada}
%   % \city{Montreal, QC}
%   % \country{Canada}
% }
% \email{{razieh.shirzadkhani, elahe.kooshafar, farimah.poursafaei }@mila.quebec, shenyang.huang@mail.mcgill.ca, rrabba@cs.mcgill.ca}

%% file: 00Abstract.tex
% This paper introduces \tgx, a Python package designed for the analysis and visualization of temporal graphs.
% Unlike existing tools which largely focus on static graphs, \tgx provides utilities for dynamic, time-evolving graphs. It offers temporal graph data loader, which is compatible with both costum and built-in datasets, as well as different statistical and visualization tools, including Temporal Edge Appearance (TEA) and Temporal Edge Traffic (TET) plots, as well as statistical metrics. 
% This package facilitates learning on temporal graphs, thus enabling the development of more accurate temporal graph learning models for various applications such as social network analysis, citation graphs and user click streams \EK{clickstream analysis? or analysis of clickstream data?}. To access \tgx and explore its capabilities, please visit the package repository at \href{https://github.com/fpour/TGX}{https://github.com/fpour/TGX}.

Real-world networks, with their evolving relations, are best captured as temporal graphs. % Real-world networks consist of relations that change over time and are often modeled as Temporal Graphs. 
However, existing software libraries are largely designed for static graphs where the dynamic nature of temporal graphs is ignored. Bridging this gap, we introduce \tgx, a Python package specially designed for analysis of temporal networks that encompasses an automated pipeline for data loading, data processing, and analysis of evolving graphs. \tgx provides access to \emph{eleven} built-in datasets and \emph{eight} external Temporal Graph Benchmark~(TGB) datasets as well as any novel datasets in the \textit{.csv} format. 
Beyond data loading, \tgx facilitates data processing functionalities such as discretization of temporal graphs and node sub-sampling to accelerate working with larger datasets. For comprehensive investigation, \tgx offers network analysis by providing a diverse set of measures, including average node degree and the evolving number of nodes and edges per timestamp.
Additionally, the package consolidates meaningful visualization plots indicating the evolution of temporal patterns, such as Temporal Edge Appearance (TEA) and Temporal Edge Traffic (TET) plots. 
The \tgx package is a robust tool for examining the features of temporal graphs and can be used in various areas like studying social networks, citation networks, and tracking user interactions. 
% The \tgx package provides a powerful tool for the study of temporal graph properties and can be utilized in many applications such as analyzing social networks, citation networks, user interaction networks, and more. 
We plan to continuously support and update \tgx based on community feedback. 
\tgx is publicly available on: \href{https://github.com/ComplexData-MILA/TGX}{https://github.com/ComplexData-MILA/TGX}.

%% file: 01Introduction.tex
\section{Introduction}

% \begin{figure}[ht]
% \centering
% % \includegraphics[width=\columnwidth]{figures/TGX_overview.pdf}
% \includegraphics[width=\columnwidth]{figures/flowchart_new.pdf}
% \caption{\tgx Overview.
% \AH{I made a simplied and updated draft, let me know what you think, might need more colors}
% \FP{Please make sure no functionality is missing.}
% } 
% % Drawing is here: https://www.mathcha.io/editor/lwx6mUlJTENfB1SlJqy3XsPMpr7PFpv5y1Zi73gwKM
% \label{fig:tgx}
% \end{figure}

\begin{figure}[t]
\centering
\includegraphics[width=0.9\columnwidth]{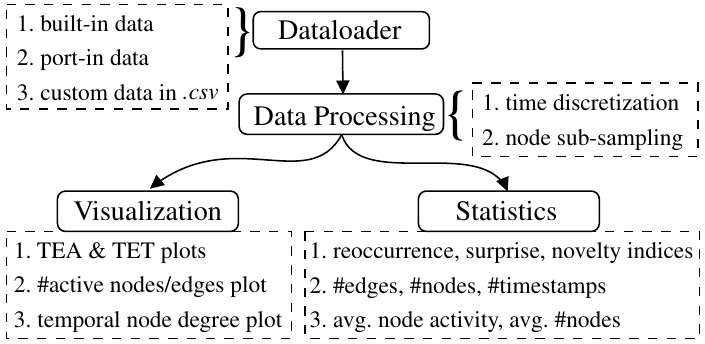}
\caption{Overview of \tgx main modules including data-loader, data processing, visualization, and statistics.}
% \FP{More space efficient version (?!).} \AH{Looks better to me.} 
% Drawing is here: https://www.mathcha.io/editor/qjNpES6qiYVIPvF5JmX6XFXddJx7ipqDMegiGqL69L
\label{fig:tgx}
\vspace{-0.21in}
\end{figure}

Recent advances in graph-based data mining has shown promising applications in different areas, such as Web data analysis, air/road traffic monitoring, and epidemic modelling. This is mostly due to the recent advancements in graph representation learning where powerful models have been developed for prediction tasks on graphs \cite{xu2018powerful}. %. \EK{Recent advances in graph-based data mining have significantly bolstered its applicability across diverse domains in web data analysis, including the analysis of social networks, citation networks, clickstream data, air/road traffic monitoring, epidemic modeling, etc. This is mostly due to the recent advancements in graph representation learning and the development of robust models designed for learning on dynamic graphs \cite{kipf2016semi, velivckovic2017graph}.}
There are also commonly used packages for training and benchmarking different graph learning models % \EK{There are several commonly used(?) packages available for training and benchmarking various graph learning models}
% [, mainly 
such as PyTorch Geometric~\cite{fey2019fast}, DGL~\cite{wang2019deep} and StellarGraph~\cite{StellarGraph}. The main library currently used for visualization and study of patterns of graphs is Networkx~\cite{hagberg2008exploring} on which DGL and PyTorch Geometric packages are dependent for statistics and visualization utilities.

Even though real-world networks typically change over time, the majority of existing literature and tools predominantly concentrate on static graphs. 
While analyzing static graphs is computationally less demanding in terms of determining statistical and algebraic properties as well as in applying machine learning models, using this static perspective results in loss of important time related features and events. For example, in epidemic modeling, the chronological order of contacts is vital in disease propagation, while static networks assume that all the contacts happen at once. %\EK{Many real-world networks can be modeled using a dynamic graph, where nodes represent entities and edges depict the relations between them. However, the prevailing body of literature and toolsets available for graph learning primarily attends to static graphs. Unlike existing tools which largely focus on static graphs, \tgx provides utilities for dynamic, time-evolving graphs. While the analysis of static graphs delivers computational simplicity, this approach overlooks pivotal time-dependent features and events in dynamic networks. For example, in epidemic modeling, the chronological order of contacts plays a vital role in disease propagation, while in static networks it is erroneously assumed that all the contacts are simultaneous. }

Therefore, there are recent efforts to extend graph learning methods to temporal graphs where the structure and attributes of graph elements are changing over time \cite{kazemi2020representation}.
Several temporal graph learning benchmarking suits have been developed for assessing TG methods, mainly PyThorch Geometric Temporal \cite{rozemberczki2021pytorch}, DyGLib \cite{yu2023towards}, and TGB \cite{huang2023temporal}. While these packages facilitate method evaluation and development with easy benchmarking, they are not essentially designed for providing comprehensive statistical analysis or visualizations for  deeper understanding of temporal graph characteristics. 
This holds significant importance in the context of model design, particularly in the incorporation of temporal patterns within datasets in addition to the comprehensive analysis of their scalability, efficiency, and interpretability. % \EK{Therefore, there has been a  surge of interest in extending static graph learning methods to temporal graphs, where the structure and attributes of graph elements are changing over time \cite{kazemi2020representation}. Several temporal graph learning benchmarking suits have been developed for assessing TG methods, mainly DyGLib \cite{yu2023towards}, BenchTemp \cite{huang2023benchtemp}, TGB \cite{huang2023temporal}, and PyThorch Geometric Temporal \cite{rozemberczki2021pytorch}. While these packages facilitate method evaluation and development by providing straightforward benchmarking tools, their primary focus is not on statistical analysis or visualization capabilities for a deeper understanding of temporal graph datasets. 
%This holds significant importance in the domain of model design, particularly with regard to studying temporal patterns in temporal graphs and conducting comprehensive analyses pertaining to their scalability, efficiency, and interpretability. }

\begin{table*}[t]
% \footnotesize
\caption{Complete list of built-in and port-in datasets in \tgx with their computed statistics.}
    \resizebox{\linewidth}{!}{%
    \begin{tabu}{l l|l r r r l l r r r r r l}
    \toprule
     & Dataset & Domain &  Nodes & Total Edges & Unique Edges & Timestamp & Duration & Unique Steps & Reoccurrence & Surprise & Node Activity & Novelty & Disc.\\ 
    \midrule
    \multirow{11}{*}{\rotatebox[origin=c]{90}{\textbf{\textcolor{blue}{Built-in}}}} &
    Reddit & Social & 10,984 & 627,447 & 78,516 & Unix & 1 month & 669,065 & 0.250 & 0.280 & 0.617 & 0.266 & daily\\
    & MOOC & Interaction & 7,144 & 411,749 & 178,443 & Unix & 1 months & 345,600 & 0.044 & 0.786 & 0.161 & 0.755 & daily\\
    & LastFM & Interaction & 1,980 & 1,293,103 & 154,993 & Unix & 4 years & 1,283,614 & 0.226 & 0.369 & 0.630 & 0.332 & monthly\\
    & Enron & Social & 184 & 125,235 & 3,125 & Unix & 3 years & 22,632 & 0.255 & 0.369 & 0.409 & 0.322 & monthly\\
    & Social Evo. & Proximity & 74 & 2,099,519 & 4,486 & Unix & 8 months & 565,932 & 0.401 & 0.027 & 0.806 & 0.127 & weekly\\
    & UCI & Social & 1,899 & 59,835 & 20,296 & Unix & 196 days & 58,911 & 0.037 & 0.796 & 0.177 & 0.651 & weekly\\
    & Flights & Transport & 13,169 & 1,927,145 & 395,072 & days & 4 months & 122 & 0.265 & 0.400 & 0.348 & 0.194 & daily\\
    & Can. Parl. & Politics & 734 & 74,478 & 51,331 & years & 14 years & 14 & 0.063 & 0.654 & 0.458 & 0.673 & yearly\\
    & US Legis. & Politics & 225 & 60,396 & 26,423 & congress & 12 cong. & 12 & 0.173 & 0.475 & 0.448 & 0.437 & yearly\\
    & UN Vote & Politics & 201 & 1,035,742 & 31,516 & years & 72 years & 72 & 0.869 & 0.027 & 0.709 & 0.056 & yearly\\
    & Contact & Proximity & 694 & 2,426,280 & 79,530 & 5 minutes & 1 month & 8,065 & 0.232 & 0.289 & 0.698 & 0.422 & daily\\
    \midrule
    \multirow{8}{*}{\rotatebox[origin=c]{90}{\textbf{\textcolor[HTML]{1b9e77}{TGB}}}} & 
    tgbl-wiki & Social & 9,227 & 157,474 & 18,257 & Unix & 1 month & 152,757 & 0.130 & 0.554 & 0.160 & 0.475 & daily\\
    & tgbl-review & Rating & 352,637 & 4,873,540 & 4,730,223 &  Unix  & 22 years & 6,865 & 0.0004 & 0.999 & 0.249 & 0.999 & yearly\\
    & tgbl-coin & Transaction & 638,486 & 22,809,486 & 3,862,031 &  Unix  & 8 months & 1,295,720 & 0.129 & 0.435 & 0.233 & 0.430 & weekly\\
    & tgbl-comment & Social & 994,790 & 44,314,507 & 35,531,704 & Unix & 5 years & 30,998,030 & 0.006 & 0.964 & 0.054 & 0.910 & monthly\\
    & tgbl-flight & Transport & 18,143 & 67,169,570 & 2,309,707 & Unix & 3 years & 1,385 & 0.295 & 0.363 & 0.652 & 0.258 & monthly\\
    & tgbn-trade & Trade & 255 & 507,497 & 34,211 & year & 32 years  & 32 & 0.702 & 0.088 & 0.868 & 0.091 & yearly\\
    & tgbn-genre & Interaction & 992 & 17,858,395 & 133,758 & Unix & 4 years & 133,758 & 0.411 & 0.246 & 0.594 & 0.189 & monthly\\
    & tgbn-reddit & Social & 11,068 & 27,174,118 & 516,669 & Unix & 3 years & 21,889,537 & 0.465 & 0.211 & 0.614 & 0.211 & monthly\\
    % \rowfont{\color{red}}
    % & tgbn-token & Transaction & 61,756 & 72,936,998 & xxx & Unix & xxx & 2,036,524 & xxx & xxx & xxx & xxx & weekly\\
        \bottomrule
        \end{tabu}
    }
    \label{tab:datasets}
\vspace{-0.1in}
\end{table*}

In this work, we introduce \tgx , a Python package designed for analysis and visualization of temporal graphs. \tgx builds upon our prior work~\cite{poursafaei2022towards}, further extending it to enhance the evaluation and analysis process for learning on temporal graphs. %\EK{Building upon our prior work~\cite{poursafaei2022towards}, \tgx aims to facilitate (foster?) a more robust evaluation and analysis process for learning on temporal graphs.} 
Here, we package the \textit{Temporal Edge Appearance (TEA)} and \textit{Temporal Edge Traffic (TET)} plots, which  are essential for investigating evolutionary patterns of temporal graph datasets. %\EK{We have packaged various tools within \tgx, including Temporal Edge Appearance (TEA) and Temporal Edge Traffic (TET) plots which provide valuable information on edge dynamics in temporal graphs and can be utilized in studying the dynamic properties in temporal graphs. }
In addition, we utilize time discretization as a means to %\EK{Furthermore, we employ time discretization as a tool to} 
control the trade-off between the computational complexity involved in generating these plots and the richness of information they offer. We also facilitate easy access to eleven commonly used small/medium-scale datasets provided in~\cite{poursafaei2022towards}, as well as  eight large-scale Temporal Graph Benchmarks (TGB)~\cite{huang2023temporal}. 
\tgx is designed to complement the benchmarking package that we recently introduced in TGB~\cite{huang2023temporal}, providing researchers with enhanced capabilities to comprehensively explore and analyze temporal patterns within these datasets and beyond.%\vspace{-0.1in}

%\EK{Main features include ...} 

%% We can sumamrize the main features, contributions as bullet points

%% file: 04Preliminaries.tex
% \vspace{-0.2in}

\section{Temporal Graph Preliminaries}

% A dynamic contact graph can be represented as a sequence of graph snapshots, $\mat{G} = \{ \mat{G}_t \}_{t=1}^{T} = \{ (\mat{V}_t, \mat{E}_t) \}_{t=1}^{T}$, where each $\mat{G}_t = ( \mat{V}_t, \mat{E}_t )$ is the graph snapshot at time $t \in [ 1 \dots T ]$. %The average node degree across all snapshots is denoted with $k$, which indicates the sparsity. The terms graph and network are used interchangeably in this paper.

% \AH{reworking this section, this this work, we deal with both continuous time dynamic graphs and discrete time dynamic graphs.} 
Temporal graphs are often categorized into Continuous Time Dynamic Graphs~(CTDGs) and Discrete Time Dynamic Graphs~(DTDGs)~\cite{kazemi2020representation}. CTDGs are considered to be the most general format of temporal graphs and can be discretized into DTDGs~\cite{souza2022provably}. In \tgx, the input is a CTDG and it supports the conversion from CTDGs into DTDGs by discretization as explained in Section~\ref{sub:discretize}. 

% Here, we define the notations for both CTDGs and DTDGs. 
CTDGs are often represented as an edge-list in the form of $\mathbf{G} = \{(u_1, v_1, t_1), (u_2, v_2, t_2), ...\} $ where edges are sorted based on timestamps, i.e., $0 \leq t_1 \leq t_2 \leq ... \leq T$ with $T$ being the maximum timestamp in the dataset. Within a certain time interval $[t_a, t_b]$, a CTDG can be discretized into a cumulative graph snapshot $\mathcal{G}_{t_a, t_b}$ constructed from all edges in the stream with timestamps $t$ where $t_a \leq t \leq t_b$  with nodes $\mathcal{V}_{t_a, t_b}$ and edges $\mathcal{E}_{t_a, t_b}$. In this way, one can evenly divide the interval $[t_1,T]$ into $k$ equal size intervals, thus creating a sequence of graph snapshots and forming the DTDG $\mathcal{G}$ discretized from the CTDG $\mathbf{G}$ where $\mathcal{G} = \{ \mathcal{G}_t \}_{t=1}^{k} = \{ (\mathcal{V}_t, \mat{E}_t) \}_{t=1}^{T}$. For clarity, we omit the $[t_a, t_b]$ subscripts when describing DTDGs as each snapshot is from an equal size interval.

% \vspace{-0.2in}

%% file: 02Features.tex
\section{Features}

\subsection{Temporal Graph Loader}
\tgx provides access to temporal graph datasets and benchmarks. Users can either load their custom datasets into the package or take advantage of built-in or ported datasets. Custom datasets can be loaded by providing the temporal edge-list of graphs, containing triplets of source, destination and timestamp. The edge-list can be loaded through the command \codeword{read_edgelist}. A complete list of currently built-in and port-in datasets and some of their important statistical properties are presented in Table~\ref{tab:datasets}. 
%The edge-list should be in the form of $\mathbf{G} = \{(u_1, v_1, t_1), (u_2, v_2, t_2), ...\} $ where edges are sorted based on timestamps, i.e., $0 \leq t_1 \leq t_2 \leq T$ with $T$ being the maximum timestamp in the dataset. 

\textbf{Built-in datasets.}
\tgx offers access to eleven built-in datasets from ~\cite{poursafaei2022towards}. 
These datasets are from different domains with various sizes and can be useful in temporal graph learning. These datasets can be directly loaded; for example, the \texttt{MOOC} dataset can be loaded via command \codeword{tgx.builtin.mooc}.

\textbf{Porting-in TGB datasets.}
The second group of temporal graph datasets that can be loaded through \tgx are the TGB~\cite{huang2023temporal} datasets. This collection includes eight temporal graph datasets from various domains and sizes, containing two medium-sized datasets with 5-25 million edges, and three large-sized datasets with over 25 million edges. 
As an example, the command \codeword{tgx.tgb_data(``tgbl-wiki")} can be used for importing the \texttt{wikipedia} dataset via TGB package.
% TGB~\cite{huang2023temporal}, an ever-evolving repository, will be continuously maintained by the Digital Research Alliance of Canada, which is funded by the Government of Canada.

\subsection{Time Discretization} \label{sub:discretize}
In temporal graphs, edges often arrive continuously with unique timestamps, potentially resulting in millions of timestamps. To effectively visualize the high-level information of a temporal graph, discretization might be necessary. 
In \tgx, users can specify their discretization preferences by specifying the desired time interval for the discretized data. Discretization intervals can be provided in two ways. The first option allows users to choose from time granularities such as specifying \codeword{``daily"}, \codeword{``weekly"}or \codeword{``monthly"}. In this case, it is assumed that the timestamps adhere to the Unix timestamp format.
Alternatively, users can provide the number of bins as an integer,  specifying the number of time slices where the entire time duration will be divided into equal time intervals. In this scenario, the timestamp unit is irrelevant. Discretization is performed directly on \codeword{tgx.Graph} object with the function \codeword{Graph.discretize}. Note that the downside of discretization is the loss of information resulting from grouping multiple edges with fine-grained timestamps into a single coarse-grained timestamp.

%Discretization can be done either during the data loading using the \codeword{read_edgelist} function by specifying the time intervals or after at any later time using \codeword{edgelist_discretizer} method.
% \vspace{-0.1in}
\subsection{Graph Sub-sampling}
In large graphs with millions of nodes, generating dataset statistics or visualization often proves challenging. 
A common approach is to sample a subset of nodes to represent the graph. In \tgx, sub-sampling can be performed on the datasets by either providing a list of nodes or specifying number of nodes to be randomly chosen by using the function \codeword{subsampling}. The sub-sampled graph will only include edges where either the source or destination node is in the selected node set. %Users should be aware that if performing the number of chosen nodes are not large enough, there might be some timestamps when no edges are sampled. 

% \vspace{-0.1in}
\subsection{Temporal Graph Statistics}

When analyzing or experimenting on temporal graph data, it is essential to have access to important statistical attributes in order to gain a better understanding and more effective use of the data. \tgx provides several tools that are helpful in the statistical analysis of temporal graphs. A discretized graph should be provided as input.

\begin{figure}[t]
\centering
\includegraphics[width=0.8\columnwidth]{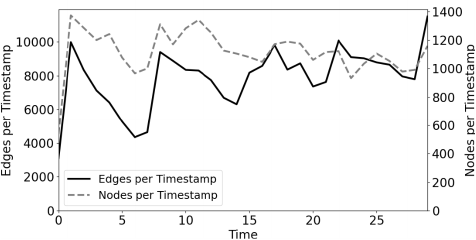}
\caption{Example plot illustrating the total number of nodes and edges per timestamp in the \texttt{MOOC} dataset discretized into 30 time intervals.}
\label{fig:nodeedge}
\vspace{-0.2in}
\end{figure}

\textbf{Total number of nodes/edges over time.}
The total number of nodes and edges over time is important in understanding the evolving nature of temporal graphs and the changes in their structure through time. For instance, Figure~\ref{fig:nodeedge} shows the total number of nodes and edges per timestamp for the \texttt{MOOC} dataset. In this visualization, the data is discretized into 30 intervals. 

% \EK{is this node activity?}
\textbf{Average node activity.}
Average node activity quantifies the average proportion of timestamps during which a node is present within a temporal network \cite{spasov2020grade}. This metric reveals the degree of temporal activity of nodes. In Table~\ref{tab:datasets}, we report the average node activity for the built-in and ported TGB datasets discretized as specified. %Average node activity for the built-in and ported TGB datasets is reported in Table~\ref{tab:datasets}. In order to calculate this metric, datasets have been discretized to the mentioned interval size in Table~\ref{tab:datasets}.  %\EK{should add reference to table 1}

\begin{figure}[h]
\centering
\includegraphics[width=\columnwidth]{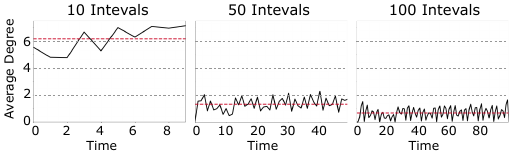}
% \caption{Example plots for the average node degree per timestamp for the MOOC dataset, discretized into three different bin counts. The red line indicates the average degree over all timestamps.} 
\caption{The average node degree per timestamp plots for the \texttt{MOOC} dataset, discretized into 10, 50 and 100 intervals. The red line indicates the average degree over all timestamps.} 
\label{fig:avedegree}
\end{figure}

\textbf{Average node degree over time.}
The visualization of the average node degree over time offers insight into the activity levels of nodes within the network. Figure~\ref{fig:avedegree} shows the average degree per timestamp in the \texttt{MOOC} dataset discretized into three different numbers of bins namely 10, 50, and 100. Notably, an increase in the number of bins corresponds to a decrease in the average degree, accompanied by more fluctuations in the trend.

\vspace{-0.2in}

\subsection{Temporal Patterns}
While temporal graphs share similarities with static graphs, conducting graph learning on temporal graphs necessitates extra metrics to ensure a robust evaluation. 
% For instance, in the context of link prediction on temporal graphs, it is essential to determine whether the predicted edges have previously existed in the graph. 
In this regard, \tgx provides access to Temporal Edge Appearance (TEA) and Temporal Edge Traffic (TET) plots as visualizations of edge dynamics as well as three metrics namely novelty, reoccurrence and surprise indices which are introduced in~\cite{poursafaei2022towards}.

\begin{figure}[t]
\centering
\includegraphics[width=0.55\columnwidth]{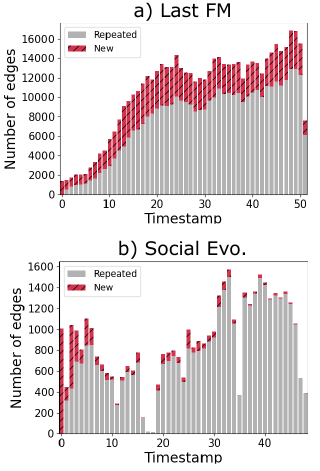}
\caption{TEA plot examples for (a) \texttt{LastFM} and (b) \texttt{Social Evo.} datasets, discretized monthly and weekly, respectively.} 
\label{fig:tea}
\vspace{-0.2in}
\end{figure}

\textbf{Temporal Edge Appearance (TEA).}
TEA plots illustrate the portion of repeated edges and newly observed edges in each timestamp. 
In \tgx, TEA plots can be generated using the command \codeword{tgx.TEA}. Figure~\ref{fig:tea} shows TEA plots for the \texttt{LastFM} and \texttt{Social Evo.} datasets discretized based on monthly and weekly timestamps, respectively. These visualizations display different temporal patterns in these datasets. As shown in Figure~\ref{fig:tea}, \texttt{LastFM} contains several new edges at each timestamp, while in \texttt{Social Evo.}, edges are mostly repeated after the first few timestamps. TEA plots can play a key role in selecting the best method for the dynamic link prediction task. Particularly, the emergence of many new edges at each timestamp results in the failure of memorization approaches. 
% Further details on TEA plots can be found in ~\cite{poursafaei2022towards}.

\begin{figure}[t]
\centering
\includegraphics[width=0.55\columnwidth]{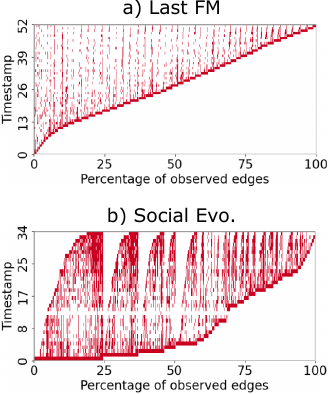}
\caption{TET plot examples for (a) \texttt{LastFM} and (b) \texttt{Social Evo.} datasets, discretized monthly and weekly respectively.} 
\label{fig:tet}
% \vspace{-0.17in}
\vspace{-0.4in}
\end{figure}

\begin{figure}[t]
\centering
\includegraphics[width=0.55\columnwidth]{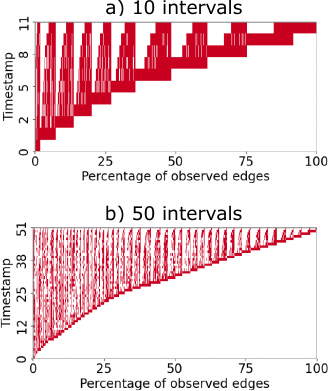}
\caption{TET plot examples for \texttt{tgbn-reddit} dataset. (a) Data discretized into 10 intervals. (b) Sub-sampled graph from 1000 random nodes, discretized into 50 intervals.} \label{fig:tet-2}
\end{figure}

\textbf{Temporal Edge Traffic (TET).}
TET plots illustrate the reoccurrence pattern of edges over time intervals. They provide more insights into the edge repetition pattern and sparsity of the network. 
TET plots are accessible through \codeword{tgx.TET} in the \tgx package. Figure~\ref{fig:tet} shows the TET plots for the \texttt{LastFM} and \texttt{Social Evo.} datasets, revealing temporal patterns in these datasets. Notably, in the \texttt{Social Evo.} dataset, a significant portion of edges repeats over time, also periods of inactivity can be observed. Figure~\ref{fig:tet-2} shows TET plots for the \texttt{tgbn-reddit} dataset. In Figure~\ref{fig:tet-2}.a, the data is discretized into 10 intervals, whereas, figure~\ref{fig:tet-2}.b displays the same dataset discretized into 50 intervals, along with graph sub-sampling involving 1000 randomly selected nodes. Both discretization and sub-sampling changes the appearance of TET plots.%[In 

\textbf{Novelty, Reoccurrence and Surprise.}
These indices were originally introduced in~\cite{poursafaei2022towards} to quantify the patterns observed in TEA and TET plots. The first index, novelty, measures the average proportion of new edges at each timestamp and is defined as \(\textit{novelty} = \frac{1}{T} \sum_{t=1}^{T} \frac{|\mat{E}^{t} \backslash \mat{E}^{t}_{seen}|}{|\mat{E}^{t}|} \), where $\mat{E}^t$ is the set of edges at timestamp $t$, and $\mat{E}^t_{seen}$ stands for all edges seen in the previous timestamps. The other two indices, reoccurrence and surprise, quantitatively assess whether edges appear repetitively or disappear after a certain time period, respectively. The formulation of these indices are as \(\textit{reoccurrence} = \frac{|\mat{E}_{train} \bigcap \mat{E}_{test} |}{|\mat{E}_{train}|}\) and \(\textit{surprise} = \frac{|\mat{E}_{test} \backslash \mat{E}_{train} |}{|\mat{E}_{test}|}\), 
% $\mat{E}_{train}$ and $\mat{E}_{test}$ 
where $\mat{E}_{train}$ and $\mat{E}_{test}$ correspond to edges present in the train and test set, respectively. These indices are reported for the built-in and ported datasets in Table~\ref{tab:datasets}. For calculation of novelty, datasets have been discretized to the mentioned interval size in Table~\ref{tab:datasets}. These measures provide a more profound insight into a network's temporal characteristics. For instance, performing link prediction on datasets with high novelty might be more complex as a simple memorization approach might not be helpful with the predictions. 
For more details on these measures, please refer to~\cite{poursafaei2022towards}.

%% file: 05Conclusion.tex
\section{Conclusion}

In summary, \tgx provides access to various tools for network analysis and visualization of temporal graphs characteristics. \tgx can be utilized in a wide range of temporal graph learning tasks, facilitating the examination of temporal patterns. More information and discussion can be found on the project \href{https://github.com/ComplexData-MILA/TGX}{\underline{Github repository}}.
%[In summary, \tgx is a package developed for analyzing and visualization of temporal graph characteristics. \tgx provides easy access to data loading tools, temporal graph statistics calculation and visualizations as well as temporal pattern visualizations. This package can be utilized in different temporal graph learning tasks. More documentation is available at the project \href{https://github.com/fpour/TGX}{Github}.]

%% file: 08Ack.tex
\section*{Acknowledgment}
This research was supported by the Canadian Institute for Advanced Research (CIFAR AI chair
program), Natural Sciences and Engineering Research Council of Canada (NSERC) Postgraduate Scholarship-Doctoral (PGS D) Award and Fonds de recherche du Québec - Nature et Technologies (FRQNT) Doctoral Award.  
We would like to thank Samsung Electronics Co., Ldt. for partially funding this research.